# INeAT: Iterative Neural Adaptive Tomography

**Bo Xiong**[a,b,*,†] , **Changqing Su**[b,†], **Zihan Lin**[c], **You Zhou**[d], **Zhaofei Yu**[a,b,e,*]
[a] School of Computer Science and Technology, Peking University, Beijing 100871, China
[b] National Engineering Research Center of Visual Technology, Peking University, Beijing, 100871, China
[c] School of Automation, Hangzhou Dianzi University, Hangzhou 310018, China
[d] Nanjing University Medical School, Nanjing 210023, China
[e] Institute for Artificial Intelligence, Peking University, Beijing 100871, China
[†] These authors contribute equally to this work.

**Abstract**. Computed Tomography (CT) with its remarkable capability for three-dimensional imaging from multiple projections, enjoys a broad range of applications in clinical diagnosis, scientific observation, and industrial detection. Neural Adaptive Tomography (NeAT) is a recently proposed 3D rendering method based on neural radiance field for CT, and it demonstrates superior performance compared to traditional methods. However, it still faces challenges when dealing with the substantial perturbations and pose shifts encountered in CT scanning processes. Here, we propose a neural rendering method for CT reconstruction, named Iterative Neural Adaptive Tomography (INeAT), which incorporates iterative posture optimization to effectively counteract the influence of posture perturbations in data, particularly in cases involving significant posture variations. Through the implementation of a posture feedback optimization strategy, INeAT iteratively refines the posture corresponding to the input images based on the reconstructed 3D volume. We demonstrate that INeAT achieves artifact-suppressed and resolution-enhanced reconstruction in scenarios with significant pose disturbances. Furthermore, we show that our INeAT maintains comparable reconstruction performance to stable-state acquisitions even using data from unstable-state acquisitions, which significantly reduces the time required for CT scanning and relaxes the stringent requirements on imaging hardware systems, underscoring its immense potential for applications in short-time and low-cost CT technology.

*Bo Xiong, E-mail: xiongbo@pku.edu.cn and *Zhaofei Yu, E-mail: yuzf12@pku.edu.cn

## 1 Introduction

Computed Tomography (CT), a non-invasive technique renowned for its ability to achieve high-resolution three-dimensional (3D) imaging, is extensively utilized in the applications of medical imaging[1,2] and plant microstructure observation[3]. Moreover, it has been gradually expanding its application into diverse fields, including materials defect detection[4,5], fluid dynamics recording[6–10] and nondestructive testing[11]. CT primarily harnesses the penetrating power of imaging waves, with X-rays being a common choice[12], although visible light waves are also utilized in certain applications[6–10], which are also named optical CT. When these waves traverse a specimen, different structures within the object exhibit varying absorption coefficients for the imaging waves. By recording these disparities from multiple perspectives for each voxel and applying subsequent reconstruction algorithms, it becomes feasible to recover the comprehensive 3D structure of the original specimen.

The primary objective of tomographic reconstruction is to recover the 3D structural information of a specimen from its two-dimensional (2D) projections acquired at multiple viewing points. In theory, with a sufficient number of viewing angles covering the Fourier space corresponding to the image function, 3D reconstruction can be achieved by the Radon inverse transform[13]. This concept underlies the classic back-projection method[14]. The subsequent methods, like filtered back-projection (FBP)[15], FDK[16] and their variants[17], have significantly enhanced the contrast of the reconstructed results by introducing filter functions. However, obtaining an adequate number of projections can be challenging in many situations due to dose limitations, sample characteristics, equipment constraints and so on[18–20]. In such cases, iterative reconstruction methods have emerged as a viable alternative[21–27], which employ an iterative process resembling the solution of linear equations. Through continuous comparisons between theoretical projections and actual measurements, these methods iteratively update and optimize until they achieve the best result. Despite significantly increased computational demands, these methods have vastly expanded the applicability of reconstruction, achieving a wide-ranging use in commercial applications[28].

Recently, various tomography reconstruction approaches based on deep learning have emerged[29–35]. Unlike traditional approaches that depend on precise physical models[20], these methods achieve remarkable reconstruction performance by training deep neural networks with extensive datasets, especially in terms of reducing imaging doses[19], mitigating imaging perspective constraints[35,36] and enhancing reconstruction quality[20]. Typically, these approaches are optimized in conjunction with traditional methods and can be categorized into three main types based on the position of network optimization: pre-processing optimization networks[37–40], post-processing optimization networks[41–43], and direct reconstruction networks[44–47]. However, they still face challenges in generalization performance, limiting their performance on unseen or diverse data[18]. Neural radiance fields (NERF)[48], as a novel paradigm for representing 3D spatial information, demonstrating remarkable reconstruction performance, are gradually being extended to tomographic reconstruction[18,49,50]. Among these methods, a self-supervised learning-based approach is employed to learn the implicit neural representation of the 3D space by inputting limited-view 2D projections and enables the synthesis of novel viewpoint images at arbitrary image resolutions. In comparison to traditional optimization methods and deep learning-based approaches, reconstruction methods based on neural radiance fields offer superior reconstruction quality without model generalization issues. Nevertheless, they typically require a substantial amount of time to train the coordinate network. To address this challenge, NeAT replaces the original coordinate network with an adaptive octree network, enabling high-quality reconstruction in just a few minutes[18].

However, during the actual data acquisition process, hardware issues with CT scanning devices, such as motor start-stop[51], slip ring wear[52], and ageing of electronic components[53], often introduce significant perturbations in the acquisition poses, such as pose fluctuations and shift. In these cases, existing reconstruction algorithms tend to suffer a significant performance decline due to the absence of appropriate pose adjustment methods, which are typically manifested as reduced resolution and increased artefacts, and existing commercial CT systems heavily rely on expensive hardware to achieve high-precision imaging[54]. Here, we propose Iterative Neural Adaptive Tomography (INeAT), a pose-iteration optimization method based on NeAT. It primarily incorporates a pose feedback module to back-project the 3D reconstruction volume from NeAT and compare them to the original input, thereby optimizing the initial input poses to enhance the final reconstruction quality. We demonstrate the resolution improvement and artefacts reduction by INeAT under varying degrees of pose disturbances via simulated data and semi-synthetic data that closely resembled real capture data, where the multiple experiments on semi-synthetic data demonstrate its capability to enhance reconstruction performance for data with different degrees of pose disturbances in actual scenarios. Furthermore, we show that our INeAT maintains comparable reconstruction performance to stable-state acquisitions even when utilizing projections from the unstable-state phase, thereby reducing the required acquisition time in practical scenarios and significantly enhancing the method's tolerance for system stability, which could highlight its significant potential for application in short-time CT technology and alleviate the demands for system precision, making it a promising candidate for widespread application in low-cost CT technology.

## 2  Methods and Results

We introduce a new neural rendering method based on iterative schemes for CT reconstruction, leveraging the benefits of NeAT while significantly enhancing robustness to pose disturbances. Proposed methods have an Iterative Neural Adaptive Tomography (INeAT) structure, which can be primarily divided into four components, illustrated in Figure 1. The first component involves the initialization of poses, as depicted in Figure 1(a). The input typically consists of a sequence of projection images collected over 360 degrees and the poses are initialized for each projection image based on a uniform sampling assumption, which may deviate from the actual acquisition. The input image sequence along with their corresponding poses are then fed into NeAT to reconstruct a coarse 3D volume, as shown in Figure 1(b). This volume could have relatively low resolution and exhibit a significant amount of reconstruction artefacts, primarily depending on the accuracy of the poses. Nonetheless, it still retains the fundamental structural information of the original object. With this limited result information, further pose optimization can be performed. As depicted in Figure 1(c), high-density reprojection of the coarse 3D volume results in a series of projection image sequences. All input images are scrutinized to locate the projection within this sequence that exhibits the highest similarity to each input image. The corresponding poses are then obtained, generating a new set of corrected pose sequences for input images, as demonstrated in Figure 1(d). Finally, this freshly obtained set of pose sequences serves as input for the pose initialization module to commence the next iteration of reconstruction.

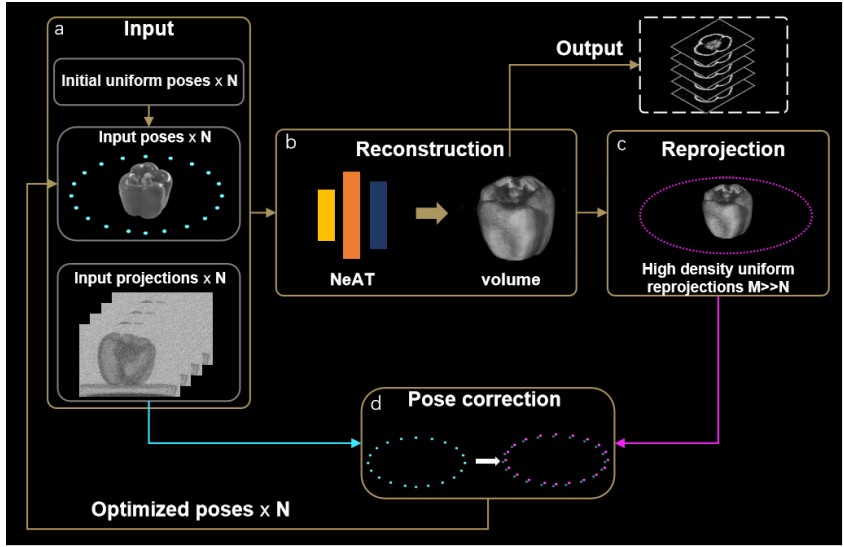

**Fig. 1** Overview of our iterative adaptive neural rendering pipeline for tomographic reconstruction. **(a)** The input module for pose initialization mainly comprises a sequence of projection images collected over 360 degrees, with the poses being initialized based on a uniform sampling assumption. **(b)** The results from (a), including projection images and their corresponding poses, are fed into NeAT to obtain the 3D implicit representation of the original object. When the desired outcome is achieved, slices in any dimension can be generated as output, and further pose optimization can be performed if necessary. **(c)** A high-density reprojection of the coarse 3D volume from (b), results in a series of projection images used for comparison in (d). **(d)** Similarity comparison between the projections from (a) and (c). This aspect enables the correction of the initial input poses by identifying the projection pairs with the highest structural similarity.

## 2.1 Design of the Iterative Neural Adaptive Tomography

Neural rendering for CT reconstruction falls under the domain of volume rendering, involving the process of transforming two-dimensional plane projections into three-dimensional volumetric densities. To ensure the accuracy of the reconstruction, it is crucial to precisely determine the correct pose information for each input projection. Due to the inaccessibility of these precise poses during data collection, we perform an initial coarse reconstruction under the assumption that these projections are strictly sampled according to a uniform distribution before we refine these poses through iterative optimization. As CT image essentially records the transmission images of a specific ray passing through an object, its imaging mode, based on the continuous version of Beer-Lambert's law[55], can be expressed as:

$$\hat{L}(p) = \hat{L}_0(p) - \int_{t_n}^{t_f} \sigma\left(r_p(t)\right) dt \tag{1}$$

where $p$ is a given image pixel, $\hat{L}_0(p)$ represents the logarithm of the intensity of the X-ray source, $r_p$ is the ray passed through the image pixel $p$, $t_n$ and $t_f$ represents the near and far intersections of the ray and the reconstruction region. By solving the 3D distribution of $\sigma(x)$, it becomes possible to determine the observed value $\hat{L}(p)$. In NeAT, it employs an optimizable explicit-implicit octree structure $\chi$ to represent the 3D space. For a ray $r_p$ generated from the position of a certain projection, it traverses the octree scene and yields a set of intersection points. The positional information and neural features of these points are subsequently fed into the decoder network $\phi$ to get the 3D distribution of $\sigma(x)$ at the corresponding locations. Typically, the ray $r_p$ is determined by the angular pose $\theta$ and the position $q = (a,b)$ in CT imaging, thus the entire process can be mathematically described as:

$$\phi + \chi : (\theta, q) \rightarrow \sigma(x) \tag{2}$$

where $q$ corresponds to the position of pixel $p$ in the projection image. To optimize both the configuration of the octree structure $\chi$ and the parameters of the decoder $\phi$, a sequence of $N$ input projections $I_s = \{I_1, I_2, \dots, I_N\}$ and a series of $N$ uniformly distributed angles $\theta_s = \{\theta_1, \theta_2, \dots, \theta_N\}$ are initialized as input data. For the angle $\theta_i, i \in \{1,2,\dots,N\}$, substituting $\theta_i$ into the equation (2) yields the 3D distribution of $\sigma(x)$. Subsequently, by incorporating $\sigma(x)$ into equation (1), we can render the observed image $\hat{I}_i$ for the current angular pose $\theta_i$. The optimization for the octree structure $\chi$ and the decoder $\phi$ can be achieved by minimizing the loss between the rendered images $\hat{I}_i$ and the original input projections $I_i$, which can be simply expressed as:

$$Loss = \sum_p \|\hat{I}_i(p) - I_i(p)\|^2 \tag{3}$$

To further enhance the quality of reconstruction, NeAT introduces additional regularization terms as constraints, as detailed in reference [18]. The entire process described above represents the complete NeAT procedure, which can be simplified as training network $\mathbb{N}$ by inputting projections $I_s$ along with their corresponding angular poses $\theta_s$ to represent the 3D distribution of $\sigma$. This can be expressed as:

$$\sigma = \mathbb{N}(I_s, \theta_s) \tag{4}$$

In real-world data, there are often varying degrees of pose disturbances in $\theta_s$. In such cases, equation (4) can only yield a rough 3D distribution of $\sigma$, which is then rendered to generate a coarse volume $V_c$. The coarse volume exhibits artefacts and inaccuracies in its structure while maintaining the fundamental shape of the reconstructed object. We aim to utilize this feature to identify the actual pose angles corresponding to the original projection. Considering that, we reproject the coarse volume $V_c$ at dense angular intervals from 0 to 360 degrees, producing the reprojection sequence $R_s = \{R_1, R_2, \dots, R_M\}$, which contains $M$ projections and their precise angle information. The procedure of reprojection could be expressed as:

$$\Gamma : V_c \rightarrow R_s \tag{5}$$

Notably, $M$ is significantly greater than the number of input projections $N$. Subsequently, we use *SSIM* (Structural Similarity Index Measure) to compare the similarity between each projection in $I_s$ and $R_s$. SSIM quantifies the structural similarity between two images. For image $I_i$ in $I_s$ and image $R_i$ in $R_s$, their structural similarity can be described mathematically as:

$$SSIM(I_i, R_i) = \frac{(2\mu_{I_i}\mu_{R_i} + c_1)(2\sigma_{I_i R_i} + c_2)}{(\mu_{I_i}^2 + \mu_{R_i}^2 + c_1)(\sigma_{I_i}^2 + \sigma_{R_i}^2 + c_2)} \tag{6}$$

where $\mu_{I_i}, \mu_{R_i}$ indicates the mean value of the image $I_i$ and $R_i$, $\sigma_{I_i}^2, \sigma_{R_i}^2$ indicates the variance of the image $I_i$ and $R_i$, $\sigma_{I_i R_i}$ is the covariance of $I_i$ and $R_i$, $c_1, c_2$ are constant values to maintain stability. As a result, for each projection in $I_s$, we find its counterpart in $R_s$ with the highest SSIM score and consider the corresponding angle as the corrected angle to replace that in $\boldsymbol{\theta}_s$. We conduct the above process on all projections in $\boldsymbol{I}_s$, obtaining a newly corrected angle sequence ($\widehat{\boldsymbol{\theta}}_s$) of the input projections:

$$SSIM(\boldsymbol{I}_s, \boldsymbol{R}_s): \boldsymbol{\theta}_s \to \widehat{\boldsymbol{\theta}}_s \tag{7}$$

Then $\boldsymbol{I}_s$ are paired with corrected poses $\widehat{\boldsymbol{\theta}}_s$ and fed into NeAT network $\mathbb{N}$ for the next iteration of reconstruction, generating the optimized reconstruction result with corrected poses, which can be expressed as:

$$\sigma_{next} = \mathbb{N}(\boldsymbol{I}_s, \widehat{\boldsymbol{\theta}}_s) \tag{8}$$

Due to the presence of structural errors and artefacts in the projections of the coarse volume, the angle correction may have limited accuracy and the results after one iteration may not make a desirable outcome. To obtain a better reconstruction result, if needed, multiple iterations involving updating the predicted angles for each input image and refining the reconstruction can be performed. We can repeat these steps until convergence is achieved, which leads to an improved and converged CT reconstruction result with optimized poses.

Additionally, the original NeAT adopted a hierarchical sampling approach based on an octree representation, employing varying resolutions in different regions of reconstruction to enhance efficiency and conserve computational resources. However, this hierarchical octree structure can result in discontinuities in resolution and structural inconsistency across nodes of different sizes. Thus we opted for a global high-resolution reconstruction strategy that train the scene with maximum resolution from the beginning. We find that employing this global high-resolution reconstruction approach not only yielded comparable reconstruction outcomes to the original NeAT, but also resulted in a time reduction of approximately 1.5 times during reconstruction to the highest resolution. This will be further elaborated in section 2.4.

*2.2 Experimental Results of the Simulation Data*

In the real-world data acquisition for CT imaging, pose uncertainty mainly results from two factors: local fluctuations due to equipment wear and tear, and acceleration-deceleration processes caused by motor start-stop. We simulated these two scenarios on synthetic data and evaluated the performance of INeAT on this dataset. Specifically, we created a three-dimensional structure where multiple cubes were arranged in a spiral pattern from bottom to top. The spiral's radius gradually increased, forming an asymmetric spiral structure on which we executed a projecting operation to generate 180 projections as input data. The maximum intensity projection tri-view of the synthesized data is shown in Figure 2(a).

For the first scenario, to simulate local fluctuations due to equipment wear and tear in CT imaging, we added a random perturbation factor to the increment of the angle between neighbouring projections to simulate the angle fluctuations during actual sampling. The pose of angle related to the *i*-th projection can be formulated as:

$$\hat{\theta}_{i+1} - \hat{\theta}_i = d + \Delta, \left(i = 1, 2, \ldots, \frac{360}{d} - 1\right) \tag{9}$$

where $\hat{\theta}_i$ represents the pose of angle corresponding to *i*-th projection and $\hat{\theta}_1$ is set as a default value of 0, *d* represents the ideal angular distance between neighbouring projections and is set as a default value of 2 in this experiment, $\Delta$ is the random perturbation factor that can be freely configured but generally less than *d* for each projection. The image depicting angles with random perturbations is illustrated in the left column of Figure 2(b), where the black dots represent default uniform input angles and the red dots represent the perturbed angles.

For the second scenario, we introduced a speed factor, which was used to alter the angular pose increment between neighbouring projections, simulating the process of acceleration and deceleration. The uniform motion phase in between still maintained constant velocity with a *d*-degree interval. According to the kinematic equation, for a given speed factor $a(>0)$, the angular pose can be described as:

$$\hat{\theta}_{i+1} - \hat{\theta}_i = \begin{cases} 0.5(ai + a(i+1)), 0 < i \leq N_a \\ 0.5(ai + d - a(i + N_a - 180)), 180 - N_a < i \leq 180 \\ d, N_a < i \leq 180 - N_a \end{cases} \quad (10)$$

where $N_a = d/a$ is the number of projections captured by the accelerating process. The diagram demonstrating this sampling method is shown in the left column of Figure 2(c).

We sampled the simulated data based on the two aforementioned angle pose sampling methods, and then obtained multiple sets of projection data by adjusting the random perturbation factor $\Delta$ and the speed factor $a$ separately. Subsequently, we tested INeAT on each of these projection datasets, as shown in Figure 2. The angular interval for reprojection in INeAT was set at 0.05 degrees, covering the entire 360-degree range and generating a total of $M = 7200$ high-density reprojected images, which were utilized for subsequent pose optimization. The ground truth image shown in "GT" column in Figure 2 is the x-y slice of the synthetic volume. It is evident that without pose optimization, the reconstruction quality degrades under the influence of the random perturbation factor, as shown in the "Default" column in Figure 2(b). Larger values of random perturbation factor $\Delta$ correspond to worse reconstruction quality. However, INeAT can significantly enhance reconstruction quality in such cases, with the improvement being more pronounced with a larger random perturbation factor, as illustrated in the "Corrected" column in Figure 2(b). This is primarily due to INeAT's ability to effectively rectify the poses of input projection images, as shown in the "Sine Curve" column in Figure 2(b). The impact of the speed factor is slightly different. A larger speed factor implies smaller pose deviations, resulting in higher reconstruction quality, while smaller factors yield worse results, as shown in the "Default" column in Figure 2(c). INeAT can likewise enhance reconstruction quality in such scenarios (the "Corrected" column in Figure 2(c)), owing to its capability to correct all poses to their corresponding actual values (the "Sine Curve" column in Figure 2(c)).

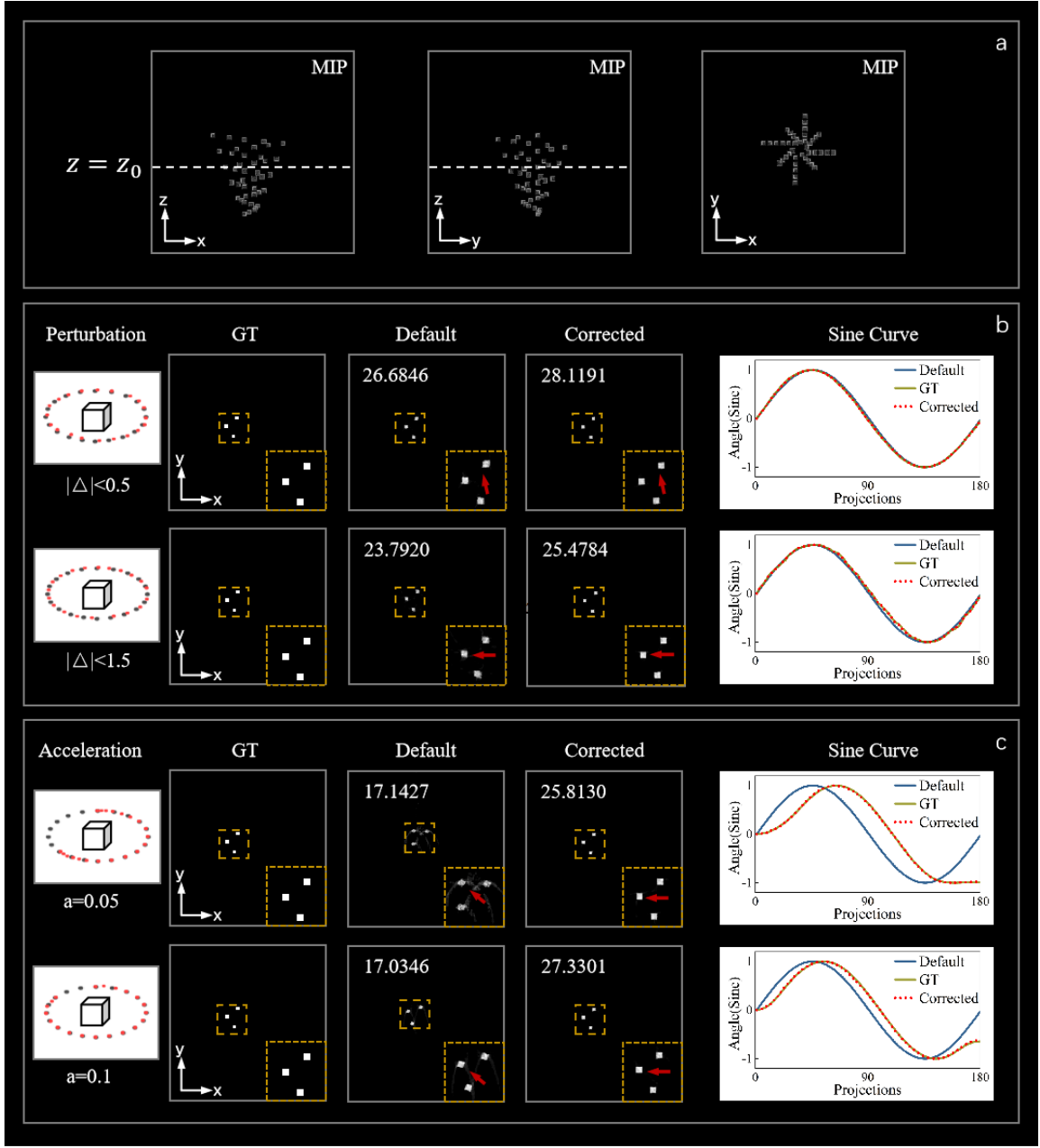

**Fig 2.** Comparison of the reconstruction results on simulated data under different scenarios of perturbation. **(a)** The maximum intensity projection tri-view of the simulated volume. The dashed line marks the z-position where the volume is sliced and displayed in (b) and (c). **(b)** Comparison of different ranges of angle perturbation added on 180 input projections and **(c)** comparison of different acceleration values under acceleration-deceleration scenes. The first columns in (b) and (c) depict schematic representations of input angles, where black dots represent the "Default" column result input angles(uniform) and red dots represent the simulation sampled angles(non-uniform) with errors. The "GT" column shows the x-y slice of the simulated volume. The "Default" and "Corrected" columns display the corresponding slice results of the reconstruction using default uniform poses and iterative corrected poses, respectively. The number on the top-left corner of the sub-image is the slice PSNR of each reconstruction to GT. The dashed box in (b) and (c) emphasizes structural details, and the red arrow highlights the regions where artefacts have been reduced and structure has been improved. The right column showcases sine curves of the pose angles, where "GT" is the accurate angle curve with perturbation in (b) or acceleration in (c), "Default" includes the default uniform angles that are fed into NeAT initially, and "Corrected" shows INeAT's corrected angles updated from the "Default".

It is worth noting that because of the slower speed during the acceleration and deceleration stage, 180 projections cannot fully cover the entire 360-degree range. However, NeAT still performs reconstruction based on the initial uniform sampling of 360 degrees, resulting in a rotation of the angle of reconstructed results from the ground truth (GT) values in the Z-axis, which exhibits a negative correlation with the speed factor, as shown in the "Default" column in Figure 2(c).

*2.3 Experimental Results of the Semi-Synthetic Data*

In real-world CT scanning scenarios, obtaining precise poses during the scanning process is typically a challenging task. To further validate the performance of our INeAT with real-world data, we introduced semi-synthetic data capable of closely mimicking the CT acquisition process in realistic scenarios. The detailed process is as follows: Initially, we employed NeAT for high-resolution reconstruction on high-precision CT scanned data, which includes 180 uniformly sampled X-ray projections and their evenly distributed angles. Within the NeAT framework, this involved increasing the depth of the octree network, with our chosen depth set at 4. This reconstruction resulted in a high-resolution 3D volume, which we regarded as the ground truth (GT). To facilitate subsequent comparisons, we standardized the resolution of this 3D volume to be isotropic, ensuring uniform resolution across all dimensions. Subsequently, we conducted data acquisition on the ground truth 3D volume, following a similar data acquisition procedure as outlined in Section 2.2 for simulated data.

In our experiment, we incorporated data from a variety of samples, including peppers[18], walnuts[56], and a rat skull[57]. We then assessed the performance of INeAT using the acquired data, which can be similarly categorized into two main types of perturbations: local fluctuations caused by equipment wear and tear, and the acceleration-deceleration process induced by the motor start-stop. Similar to Section 2.2, the degree of perturbation in both scenarios is represented by random perturbation factors $\Delta$ and speed factors $a$, and their respective impacts on the reconstruction performance of INeAT are shown in Figure 3 and Figure 4, respectively. This approach enabled us to evaluate the performance of INeAT under various degrees of pose perturbations for real-world objects.

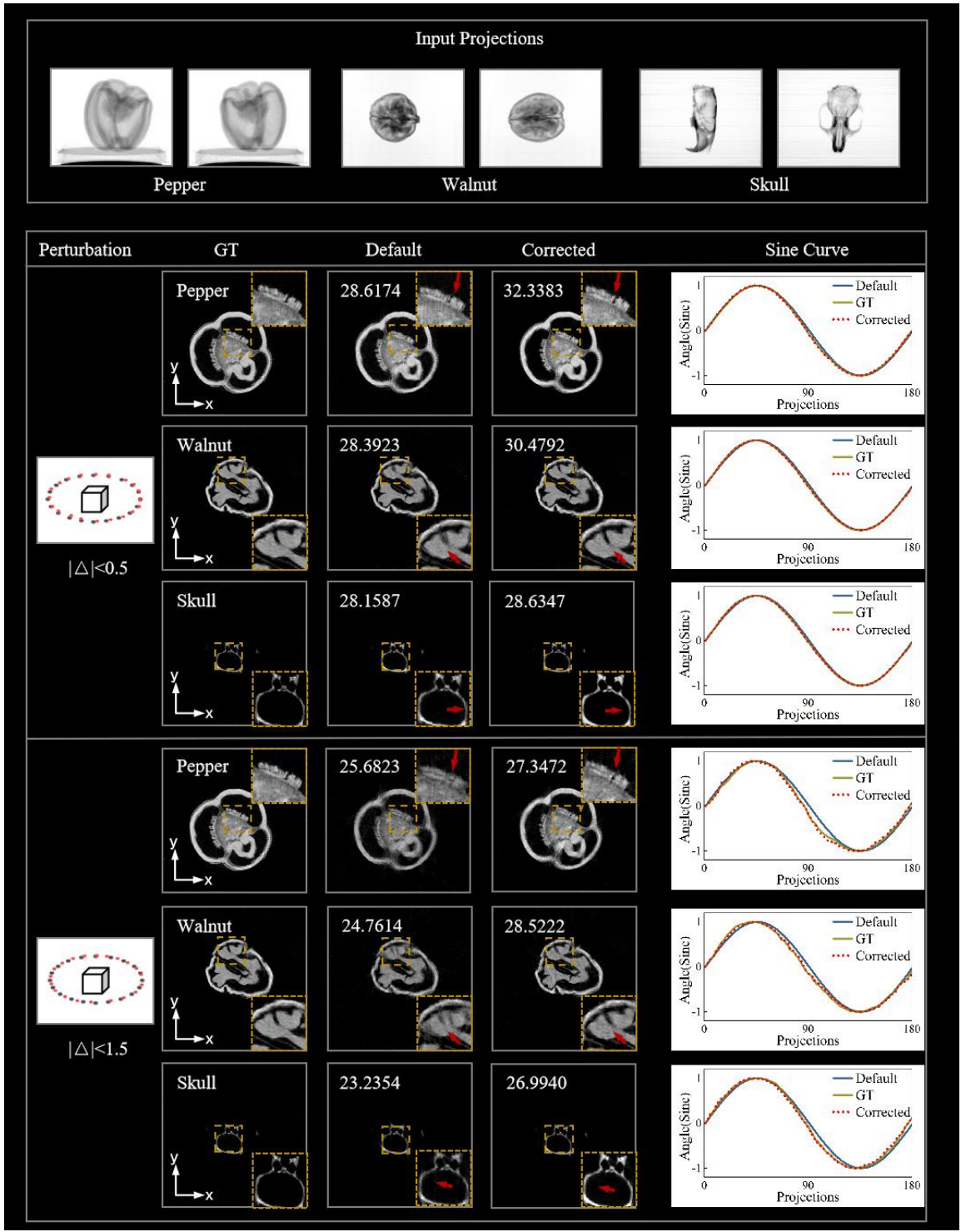

**Fig 3.** Comparison of the reconstruction results under different random perturbation factors on semi-synthetic data consisting of pepper, walnut and rat skull. The images on top are the projections of these three datasets from different angles of view. The left column shows the range of angle perturbation applied to a single projection view and the graphic illustration of the sampled angles(red dots, non-uniform) and "Default" column result input angles(black dots, uniform). The "GT" column shows the x-y slice of the GT volume of the semi-synthetic data. The "Default" and "Corrected"

columns display the corresponding slice results of the reconstruction using default uniform poses and iterative corrected poses, respectively. The number on the top-left corner of the sub-image is the slice PSNR of each reconstruction to GT. The dashed box in each sub-image emphasizes structural details, and the red arrow highlights the regions where artefacts have been reduced and structure has been improved. The right column showcases sine curves of the input angles, where "GT" is the accurate angle curve with perturbation, "Default" includes the default uniform angles that are fed into NeAT initially, and "Corrected" shows INeAT's corrected angles updated from the "Default".

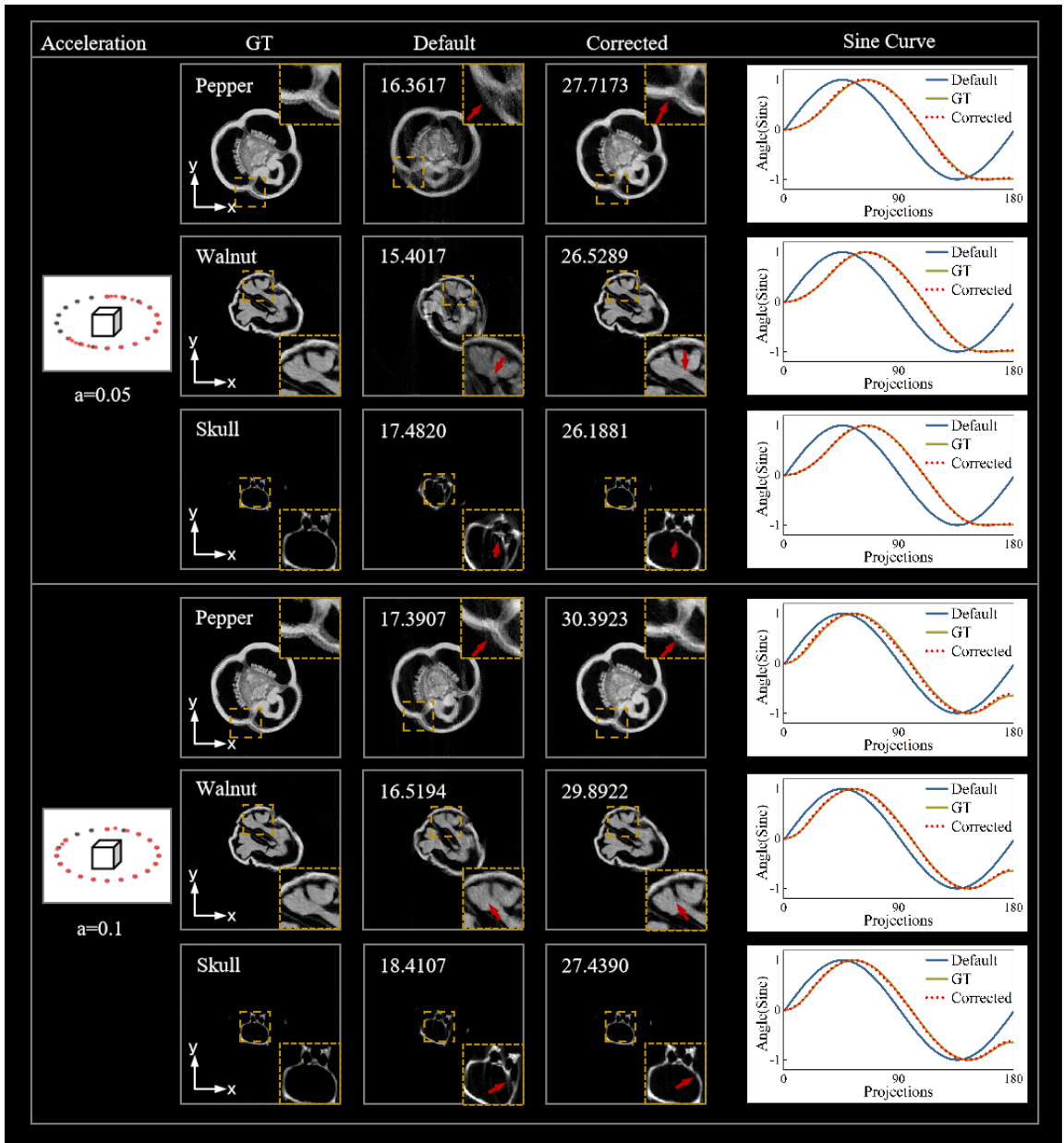

**Fig 4.** Comparison of the reconstruction results with different speed factors on semi-synthetic data consisting of pepper, walnut and rat skull. The left column shows the values of the acceleration that is applied to the accelerate-decelerate motion phase of the data acquisition and its schematic diagram of the sampled angles(red dots, non-uniform) and "Default" column result input angles(black dots, uniform). The "GT" column is the x-y slice of the GT volume of the semi-synthetic data. The "Default" and "Corrected" columns display the corresponding slice results of the reconstruction using default

uniform poses and iterative corrected poses, respectively. The number on the top-left corner of the sub-image is the slice PSNR of each reconstruction to GT. The dashed box in each sub-image emphasizes structural details, and the red arrow highlights the regions where artefacts have been reduced and structure has been improved. The right column showcases sine curves of the input angles, where "GT" is the accurate angle curve with acceleration, "Default" includes the default uniform angles that are fed into NeAT initially, and "Corrected" shows INeAT's corrected angles updated from the "Default".

We demonstrate the impact of random perturbation factors $\Delta$ on reconstruction performance in Figure 3. The random perturbation factor $\Delta$ primarily induces deviations in every single acquisition. Due to the error accumulation effect, these deviations gradually accumulate as the number of acquisitions increases. As the sampling progresses, due to the cumulative effect of the perturbation, the later angles may deviate significantly from the default angles, as shown in the GT curve in "Sine Curve" in Figure 3. When $\Delta$ is relatively small, NeAT's built-in pose correction module can make minor adjustments, ensuring that the reconstruction quality does not significantly deteriorate. As $\Delta$ increases, the pose deviation surpasses NeAT's correction capacity, leading to a noticeable decrease in reconstruction quality, as illustrated in the "Default" column in Figure 3. On the contrary, the larger the $\Delta$ value, the more substantial the relative enhancement achieved by our INeAT (in the "Corrected" column in Figure 3). This is mainly because a larger $\Delta$ value leads to more significant accumulated pose deviations during the acquisition process (in the "Sine Curve" column in Figure 3). However, the pose correction module in INeAT effectively mitigates the influence of these pose deviations in the "Sine Curve" column in Figure (3), resulting in improved reconstruction results(in the "Corrected" column in Figure 3). Moreover, the improvement becomes more pronounced as $\Delta$ increases.

The impact of speed factors $a$ on reconstruction performance is shown in Figure 4. The speed factor primarily introduces a directional pose deviation during the motor start-stop, as depicted in the "Sine Curve" column in Figure 4. A smaller speed factor corresponds to a greater pose deviation, resulting in poorer reconstruction results under the assumption of uniform sampling. (in the "Default" column in Figure 4). Additionally, due to the presence of the speed factor, the entire sampling process doesn't cover the full 360 degrees. As a result, under the default assumption of uniform sampling, there will be a rotation in the reconstruction results compared to the ground truth values around the Z-axis direction (in the "Default" column in Figure 4). However, INeAT significantly enhances reconstruction quality in such scenarios due to its robust pose correction capabilities, as shown in the "Corrected" and "Sine Curve" columns in Figure 4.

We also present the three-dimensional volume renderings of the reconstruction results in Figure 5. Figure 5(a) corresponds to the default reconstruction of NeAT with a sampling method featuring an acceleration of 0.05, while Figure 5(b) illustrates the corresponding reconstruction of INeAT with angle correction. As observed in Figure 5, due to angle errors, there were artefacts and blurriness in the three-dimensional structure of the default reconstruction. However, after angle correction, these artefacts are mitigated and structural details are restored, resulting in a significant improvement in reconstruction quality. Experiments with other values of random perturbation factor, speed factors and the combination of both will be detailed in the supplementary materials.

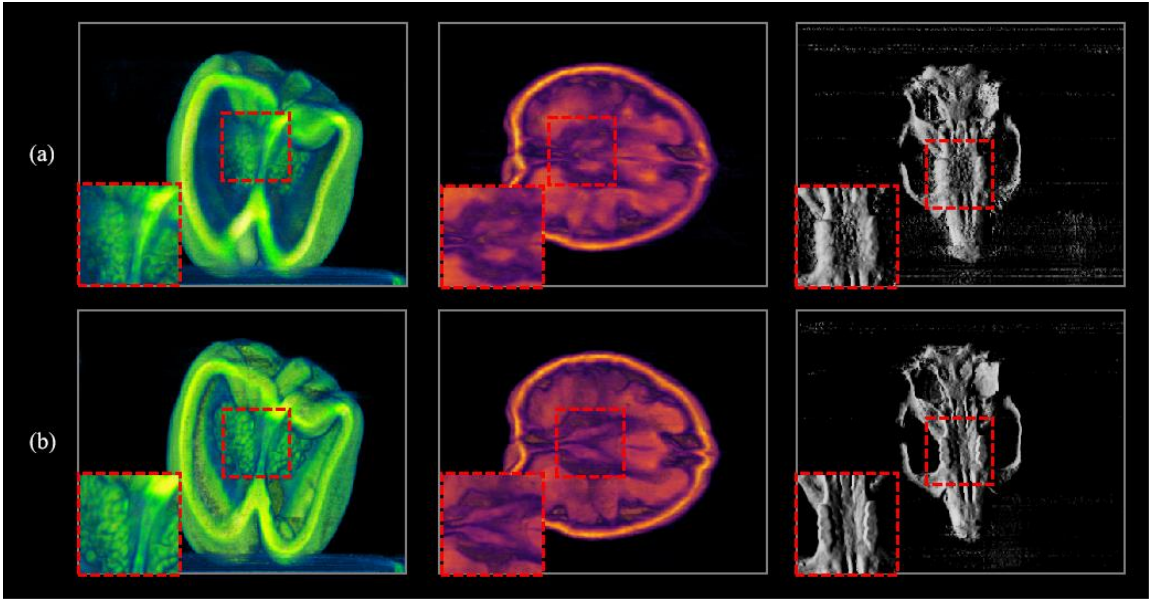

**Fig 5.** Comparison of the 3D volume renderings of semi-synthetic data consisting of pepper, walnut and skull, by NeAT and INeAT. **(a)** The default reconstruction results of NeAT under the acceleration-deceleration scene, with the speed factor $a$ set to 0.05, **(b)** The iterative reconstruction results of INeAT with angle correction applied based on (a). The structural details are emphasized in red blocks.

*2.4 A High-Resolution Alternative of the Octree Configuration*

As detailed in section 2.1, NeAT performs reconstruction by training an adaptive octree scene that can flexibly represent structures in various regions with different resolutions. Specifically, the octree is initialized with 4x4x4 blocks. As the training progresses, these blocks can be subdivided based on the information content at their respective locations, with the potential for a maximum division of 16x16x16. The blocks containing more high-frequency information have a higher likelihood of being subdivided. However, when blocks of different sizes are adjacent, such hierarchical representation can result in overly distinct boundaries, inconsistent resolution across different regions, and incompleteness in the reconstructed structure. These imperfections could affect the accuracy of the reconstruction results.

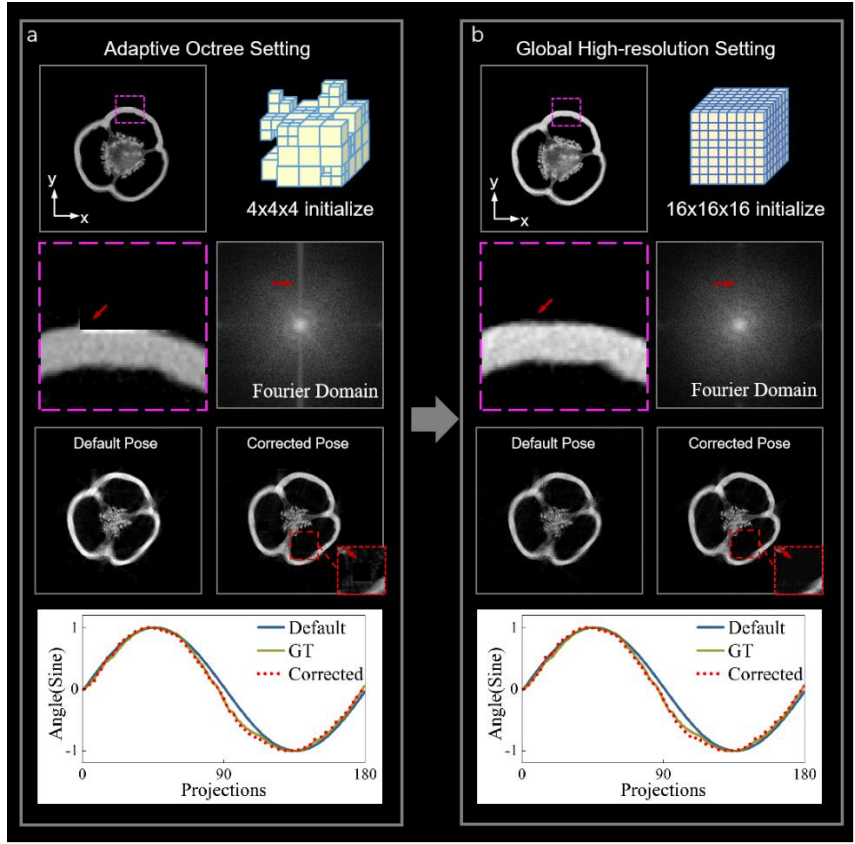

**Fig 6.** Comparison of the reconstruction result of semi-synthetic pepper data using default adaptive octree setting and adjusted global high-resolution octree setting. **(a)** The results of the default octree setting, which trains the octree from 4x4x4 and performs subdivision and cutting as reconstruction processes. **(b)** The modified octree configuration is applied to our INeAT, which trains the octree directly from 16x16x16 to obtain globally high reconstructed resolution. The first row demonstrates the reconstructed slice with precise uniform input angles and the post-train octree structure of the two different settings. The second row shows more information on the reconstructed slice, with the left picture detailing the boundary artefacts and structural absence, and the right one displaying its representation in the frequency domain using Fourier transform. The red arrows emphasize the boundary of the reconstructed structure and the corresponding signal in the frequency domain. The third row shows the iterative reconstruction results of INeAT on perturbed input data using two different octree settings, with the regions of block-like artefacts highlighted in red dashed boxes. The last row shows the corresponding sine curves of the reconstruction using the above two different octree settings.

To mitigate these issues, we adopted an alternative approach by initializing the octree with 16x16x16 blocks right from the start. This means conducting reconstruction at the highest resolution globally, which is illustrated in Figure 6. The reconstruction results using the default adaptive octree configuration exhibit noticeable artefacts, as shown in Figure 6(a). In the region bordered by the pink box, the octree block is automatically pruned due to the lack of density information on the outer side of the pepper object, while its neighbouring blocks remain preserved, resulting in incomplete structural representation in this area. Contrarily, our modified octree configuration demonstrates notable improvements in addressing the structural absence deficiencies and boundary artefacts, as shown in Figure 6(b). This is because the new configuration ensures uniform sampling across the entire region of reconstruction, avoiding abrupt changes in resolution and preventing the removal of blocks with less information. It can also be observed from the frequency domain that the modified configuration reduces the presence of block-like artefacts in the frequency domain, indicating that this configuration is effective in easing the boundary artefacts in the reconstruction results.

We also conducted reconstructions of perturbed data using these two octree settings, and the results are shown in the third and last rows in Figure 6. Both sets of data include 180 projections with <1.5-degree angular perturbation added to their poses. It can be observed that the reconstruction results with the default octree setting exhibit block omissions. This stems from NeAT's octree dividing configuration which determines whether to split or merge octree blocks based on the density of the features within that region. If features are inadequate, the block may be pruned, resulting in discontinuous block-boundary artefacts. After modifying this setting, such artefacts vanish and the entire reconstruction structure becomes more cohesive.

Furthermore, upon analyzing the angle estimation outcomes presented in the last row of the figure, there exists no substantial difference or improvements in the angles observed between the two scenarios. However, the ultimate goal of our work is to improve the quality of CT reconstruction and faithfully reconstruct the true structure of the object. So we choose to adopt the global high-resolution octree setting in INeAT. Additionally, under the same input conditions of 180 projections, the reconstruction speed using this new octree configuration is approximately 20 minutes, which is about 1.5 times faster compared to the 30 minutes using the default adaptive octree configuration.

*2.5 The potential for applications in short collection time and low-cost CT technology*

In the conventional CT scanning process, data acquisition does not typically work during the motor acceleration and deceleration phases, or data collected during these phases is usually discarded. That means only data collected during the stable velocity phase is used for reconstruction, potentially resulting in extra time in the data acquisition process. However, with our pose correction method, we can make use of the data acquired during the unstable motion phase. As shown in Figure 7, we compared the reconstruction results of the conventional sampling method and our method under the condition of an equal amount (180) of input projections. We conducted three sets of experiments: the first set (Figure 7(a)) used data obtained from 360-degree uniform sampling as input, with known and accurate angle information; This was designed to simulate the sampling of the stable-motion phase, following the same generation method as previously described for the GT. The second set (Figure 7(b)) employed 180 projections from a sampling process with both acceleration-deceleration and random perturbations for reconstruction, where "Default" represents the default NeAT reconstruction result, while "Ours" corresponds to the reconstruction result with our INeAT. The third set (Figure 7(c)) is similar to the second set, but the overall perturbations are larger compared to the second set. As demonstrated in Figure 7, our INeAT can achieve comparable reconstruction quality to that of stable velocity data even when there are acceleration and deceleration phases as well as perturbation influence in the data collection. This capability has the potential to dramatically reduce CT scanning time while simultaneously easing the demanding prerequisites for imaging hardware systems. This, in turn, could drive the development of short-time and low-cost CT technology.

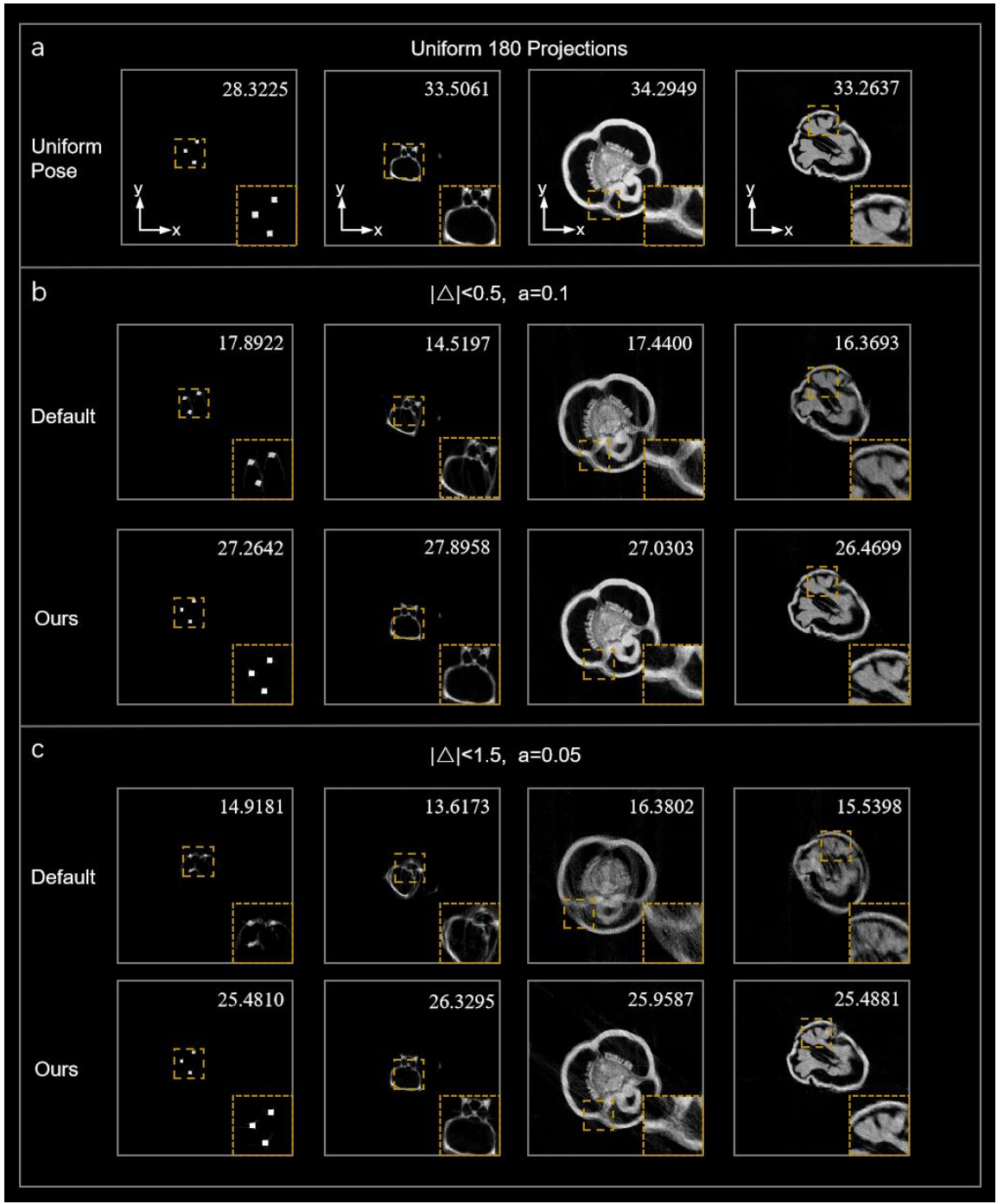

**Fig 7**. Comparison of the reconstruction results on different datasets using uniform sampling and our pose correction method with an equal amount (180) of input projections. The first column shows the results of simulated data, and the right three columns are results of rat skull, pepper and walnut respectively. **(a)** The reconstruction results using 180 projections sampled from 360 degrees uniformly with their actual angles. **(b)** The default NeAT and our INeAT reconstruction results with sampling method influenced by the joint impact of the perturbation and speed factor, where $|\Delta| < 0.5, a = 0.1$. **(c)** The default NeAT and our INeAT reconstruction results with $|\Delta| < 1.5, a = 0.05$. The dashed box in each sub-image is the magnified picture of the structure to display the reconstruction details.

# 3   Conclusion

In this paper, we introduce INeAT, an extension of NeAT that incorporates pose feedback correction. INeAT effectively corrects pose perturbations caused by motor start-stop, slip ring wear, hardware ageing, and other factors encountered during actual CT scanning processes. This correction significantly enhances reconstruction quality in these unstable scenarios. We first validate the feasibility of INeAT through experiments with simulated data. Subsequently, we demonstrate its ability to effectively transfer to real data by evaluating its impact on semi-synthetic data, leading to improved reconstruction quality in actual data. Through parameter optimization, we enhance INeAT's speed by 1.5 times while maintaining or even improving reconstruction quality. Finally, by comparing the reconstruction results of INeAT using data from unstable states with those using stable acquisition data across multiple samples, we demonstrated that INeAT can achieve reconstruction quality comparable to stable acquisition data, even with unstable data. This significantly reduces the duration of CT scanning and alleviates the exacting demands on imaging hardware systems positioning it as a promising tool for short-time and low-cost CT technology.

Since the improvement of the reconstruction quality in INeAT primarily relies on pose correction accuracy which is mainly determined by the density of reprojections, INeAT treats the density of reprojections as a parameter that can be appropriately adjusted for different scenarios. Additionally, this parameter also affects the overall reconstruction time, which may require a trade-off in some cases. Further improvement could involve combining more advanced pose correction strategies and optimizing the time consumption of reconstruction render. We firmly believe that this pose correction capability will provide the broader CT community with a high-performance tool, easing the demand for high-precision acquisition and promoting the development of cost-effective and fast CT technology.

*Supporting Information*

Supporting Information is available from the Wiley Online Library or from the author.

*Acknowledgments*

The work was supported by the National Natural Science Foundation of China (NSFC) (Grant Nos. 62371006 and 62071219) and China Postdoctoral Science Foundation (Grant No. 2023TQ0006). B. Xiong and Z. Yu supervised this project. B. Xiong and C. Su conceived and designed the experiments. C. Su and Z. Lin conducted the experiments. B. Xiong, Y. Zhou, C. Su and Z. Lin realized the neural networks. All authors discussed the results and contributed to the final manuscript.

*Conflict of interest*

The authors declare no conflicts of interest.

*Data Availability Statement*

The data and codes that support the figures and findings within this article are available from the corresponding authors upon reasonable request.

*Keywords*

Computed tomography, Implicit neural representation, Iterative posture optimization, Volume rendering

# Reference

[1] S. D. Rawson, J. Maksimcuka, P. J. Withers, S. H. Cartmell, *BMC Biol.* **2020**, *18*, 21.
[2] B. Van Ginneken, B. M. Ter Haar Romeny, M. A. Viergever, *IEEE Trans. Med. Imaging* **2001**, *20*, 1228.
[3] A. Piovesan, V. Vancauwenberghe, T. Van De Looverbosch, P. Verboven, B. Nicolaï, *Trends Plant Sci.* **2021**, *26*, 1171.
[4] S. Brisard, M. Serdar, P. J. M. Monteiro, *Cem. Concr. Res.* **2020**, *128*, 105824.
[5] L. Vásárhelyi, Z. Kónya, Á. Kukovecz, R. Vajtai, *Mater. Today Adv.* **2020**, *8*, 100084.
[6] B. Atcheson, I. Ihrke, W. Heidrich, A. Tevs, D. Bradley, M. Magnor, H.-P. Seidel, *ACM Trans. Graph.* **2008**, *27*, 132:1.
[7] M.-L. Eckert, K. Um, N. Thuerey, *ACM Trans. Graph.* **2019**, *38*, 1.
[8] J. Gregson, M. Krimerman, M. B. Hullin, W. Heidrich, *ACM Trans. Graph.* **2012**, *31*, 52:1.
[9] S. W. Hasinoff, K. N. Kutulakos, *IEEE Trans. Pattern Anal. Mach. Intell.* **2007**, *29*, 870.
[10] G. Zang, R. Idoughi, C. Wang, A. Bennett, J. Du, S. Skeen, W. L. Roberts, P. Wonka, W. Heidrich, **2020**, pp. 1870–1879.
[11] I. Amenabar, A. Mendikute, A. López-Arraiza, M. Lizaranzu, J. Aurrekoetxea, *Compos. Part B Eng.* **2011**, *42*, 1298.
[12] A. C. Kak, M. Slaney, **1988**.
[13] J. Radon, *IEEE Trans. Med. Imaging* **1986**, *5*, 170.
[14] Y. Long, J. A. Fessler, J. M. Balter, *IEEE Trans. Med. Imaging* **2010**, *29*, 1839.
[15] S. Singh, M. K. Kalra, J. Hsieh, P. E. Licato, S. Do, H. H. Pien, M. A. Blake, *Radiology* **2010**, *257*, 373.
[16] L. A. Feldkamp, L. C. Davis, J. W. Kress, *JOSA A* **1984**, *1*, 612.
[17] X. Pan, E. Y. Sidky, M. Vannier, *Inverse Probl.* **2009**, *25*, 123009.
[18] D. Rückert, Y. Wang, R. Li, R. Idoughi, W. Heidrich, *ACM Trans. Graph.* **2022**, *41*, 55:1.
[19] W. Xia, H. Shan, G. Wang, Y. Zhang, *IEEE Signal Process. Mag.* **2023**, *40*, 89.
[20] G. Wang, J. C. Ye, B. De Man, *Nat. Mach. Intell.* **2020**, *2*, 737.
[21] K. Abujbara, R. Idoughi, W. Heidrich, in *2021 Int. Conf. 3D Vis. 3DV*, **2021**, pp. 175–185.
[22] J. Huang, Y. Zhang, J. Ma, D. Zeng, Z. Bian, S. Niu, Q. Feng, Z. Liang, W. Chen, *PLOS ONE* **2013**, *8*, e79709.
[23] Y. Huang, O. Taubmann, X. Huang, V. Haase, G. Lauritsch, A. Maier, *IEEE Trans. Radiat. Plasma Med. Sci.* **2018**, *2*, 307.
[24] S. J. Kisner, E. Haneda, C. A. Bouman, S. Skatter, M. Kourinny, S. Bedford, **n.d.**
[25] E. Y. Sidky, X. Pan, *Phys. Med. Biol.* **2008**, *53*, 4777.
[26] M. Xu, D. Hu, F. Luo, F. Liu, S. Wang, W. Wu, *IEEE Trans. Radiat. Plasma Med. Sci.* **2021**, *5*, 78.
[27] G. Zang, M. Aly, R. Idoughi, P. Wonka, W. Heidrich, **2018**, pp. 137–153.
[28] H. Shan, A. Padole, F. Homayounieh, U. Kruger, R. D. Khera, C. Nitiwarangkul, M. K. Kalra, G. Wang, *Nat. Mach. Intell.* **2019**, *1*, 269.
[29] R. Cierniak, *Int. J. Appl. Math. Comput. Sci.* **2008**, *18*, 147.
[30] T. Würfl, F. C. Ghesu, V. Christlein, A. Maier, in *Med. Image Comput. Comput.-Assist. Interv. - MICCAI 2016* (Eds.: S. Ourselin, L. Joskowicz, M. R. Sabuncu, G. Unal, W. Wells), Springer International Publishing, Cham, **2016**, pp. 432–440.
[31] Y. S. Han, J. Yoo, J. C. Ye, **2016**, DOI 10.48550/arXiv.1611.06391.
[32] J. Gu, J. C. Ye, **2017**, DOI 10.48550/arXiv.1703.01382.

[33] K. H. Jin, M. T. McCann, E. Froustey, M. Unser, *IEEE Trans. Image Process.* **2017**, *26*, 4509.
[34] D. M. Pelt, J. A. Sethian, *Proc. Natl. Acad. Sci.* **2018**, *115*, 254.
[35] J. Dong, J. Fu, Z. He, *PLOS ONE* **2019**, *14*, e0224426.
[36] L. Shen, W. Zhao, L. Xing, *Nat. Biomed. Eng.* **2019**, *3*, 880.
[37] R. Anirudh, H. Kim, J. J. Thiagarajan, K. A. Mohan, K. Champley, T. Bremer, in *2018 IEEECVF Conf. Comput. Vis. Pattern Recognit.*, **2018**, pp. 6343–6352.
[38] M. U. Ghani, W. C. Karl, in *2018 IEEE 13th Image Video Multidimens. Signal Process. Workshop IVMSP*, **2018**, pp. 1–5.
[39] C. Tang, W. Zhang, Z. Li, A. Cai, L. Wang, L. Li, N. Liang, B. Yan, in *15th Int. Meet. Fully Three-Dimens. Image Reconstr. Radiol. Nucl. Med.*, SPIE, **2019**, pp. 537–541.
[40] S. Yoo, X. Yang, M. Wolfman, D. Gursoy, A. K. Katsaggelos, in *2019 IEEE Int. Conf. Image Process. ICIP*, **2019**, pp. 1252–1256.
[41] Z. Liu, T. Bicer, R. Kettimuthu, D. Gursoy, F. D. Carlo, I. Foster, *JOSA A* **2020**, *37*, 422.
[42] A. Lucas, M. Iliadis, R. Molina, A. K. Katsaggelos, *IEEE Signal Process. Mag.* **2018**, *35*, 20.
[43] D. M. Pelt, K. J. Batenburg, J. A. Sethian, *J. Imaging* **2018**, *4*, 128.
[44] J. Adler, O. Öktem, *IEEE Trans. Med. Imaging* **2018**, *37*, 1322.
[45] H. Chen, Y. Zhang, Y. Chen, J. Zhang, W. Zhang, H. Sun, Y. Lv, P. Liao, J. Zhou, G. Wang, *IEEE Trans. Med. Imaging* **2018**, *37*, 1333.
[46] J. He, Y. Wang, J. Ma, *IEEE Trans. Med. Imaging* **2020**, *39*, 2076.
[47] E. Kang, W. Chang, J. Yoo, J. C. Ye, *IEEE Trans. Med. Imaging* **2018**, *37*, 1358.
[48] B. Mildenhall, P. P. Srinivasan, M. Tancik, J. T. Barron, R. Ramamoorthi, R. Ng, *Commun. ACM* **2021**, *65*, 99.
[49] Y. Sun, J. Liu, M. Xie, B. Wohlberg, U. S. Kamilov, *IEEE Trans. Comput. Imaging* **2021**, *7*, 1400.
[50] G. Zang, R. Idoughi, R. Li, P. Wonka, W. Heidrich, **2021**, pp. 1960–1970.
[51] J. Hsieh, *Computed Tomography: Principles, Design, Artifacts, and Recent Advances*, SPIE Press, **2003**.
[52] N. FUKUDA, K. SAWA, T. UENO, in *2018 IEEE Holm Conf. Electr. Contacts*, **2018**, pp. 521–525.
[53] M. Alam, H. Shen, N. Asadizanjani, M. Tehranipoor, D. Forte, *IEEE Trans. Device Mater. Reliab.* **2017**, *17*, 59.
[54] A. T. Susanto, P. Prajitno, K. Kurnianto, *J. Phys. Conf. Ser.* **2021**, *1825*, 012033.
[55] A. C. Kak, M. Slaney, *Principles of Computerized Tomographic Imaging*, Society For Industrial And Applied Mathematics, **2001**.
[56] M. J. Lagerwerf, S. B. Coban, K. J. Batenburg, **2020**.
[57] S. B. Coban, **2017**, DOI 10.5281/zenodo.1164088.